\documentclass[letter,reprint,amsmath,amssymb,prl]{revtex4-2}
\usepackage{graphicx}% Include figure files
\usepackage{dcolumn}% Align table columns on decimal point
\usepackage{bm}% bold math
\usepackage{hyperref}% add hypertext capabilities
\usepackage{tikz}
\usepackage{float}
\usepackage{verbatim}
\usepackage{booktabs}% use /midrule instead of /hline to better vertically space the line below a table header

\begin{document}

\title{High Resolution Study of \texorpdfstring{$^{40}\mathrm{Ca}$}{40Ca} to Constrain Potassium Nucleosynthesis in NGC 2419}

\author{W. Fox$^{1,2}$, R. Longland$^{1,2}$, C. Marshall$^{2,3}$, and F. Portillo Chaves$^{1}$}

\affiliation{%
 $^{1}$Department of Physics, North Carolina State University, Raleigh, North Carolina 27695
}%
\affiliation{%
 $^{2}$Triangle Universities Nuclear Laboratory, Durham, North Carolina 27708
}%
\affiliation{%
 $^{3}$Department of Physics $\&$ Astronomy, University of North Carolina, Chapel Hill, NC 27599
}%

\date{\today}

\begin{abstract}
The globular cluster NGC 2419 was the first to exhibit a Mg--K anticorrelation, linked to hydrogen burning at temperatures between 80--260 MK. However, the key K-destroying reaction, $^{39}\mathrm{K}(p,\gamma)^{40}\mathrm{Ca}$, has a large rate uncertainty in this range. We significantly constrain this rate with a high resolution $^{39}\mathrm{K}(^{3}\mathrm{He},d)^{40}\mathrm{Ca}$ study. We resolve the E$_{\text{r}}^{\text{c.m.}} = 154$ keV resonance in $^{39}\mathrm{K}+p$ for the first time, increasing the previous rate by up to a factor 13 and reducing its $1\sigma$ width by up to a factor of 42. Reaction network calculations for NGC 2419 suggest that this could lower temperatures needed to reproduce the Mg--K anticorrelation.
\end{abstract}

\maketitle

%%%%%%%%%%%%%%%%%%%%%%%%%%%%%%%%%%%%%%%%%%%%%%%%%%%%%%%%%%%%%%%%%%%%%%%%%%%%%%%%%%%%%%%%
%%%%%%%%%%%%%%%%%%%%%%%%%%%%%%%%%%%%%%%%%%%%%%%%%%%%%%%%%%%%%%%%%%%%%%%%%%%%%%%%%%%%%%%%
%\section{Introduction}

Globular cluster evolution is poorly understood, despite decades of research. The discoveries of multiple stellar populations \cite{Bedin2004,Piotto2007} and abundance anticorrelations between light element pairs, such as O--Na and Mg--Al (see Gratton \emph{et al.} \cite{Gratton2019} for a recent review), point to a more complex history than previously assumed. Abundance anticorrelations, in particular, are exhibited between low-mass globular cluster stars and therefore cannot have originated through nuclear burning \emph{in situ}. These stars do not produce sufficient temperatures to account for changes in abundance of the element pairs \cite{Prantzos2007}. The anticorrelations likely stem from mixing with the matter of older stars in the cluster, called polluter stars.

The discovery of a Mg--K anticorrelation among red giant stars in globular cluster NGC 2419 \cite{Mucciarelli2012,Cohen2012} presented another puzzle in the effort to understand globular cluster evolution (these patterns have since been observed in NGC 2808 and $\omega$ Centauri \cite{AlvarezGaray2022}). About $30\%$ of the red giants observed in NGC 2419 exhibit a strong potassium enhancement, corresponding with a considerable magnesium depletion, indicating that these stars went through an unknown process. Ventura \emph{et al.} \cite{Ventura2012} were the first to propose polluter star candidates for the Mg--K anticorrelation in NGC 2419. Hot-bottom burning in massive asymptotic giant branch (AGB) stars and super-AGB (SAGB) stars was considered, where sufficient potassium production and magnesium depletion can be achieved, but only in a specific configuration of stellar model parameters.

Iliadis \emph{et al.} \cite{Iliadis2016} expanded the search for polluter candidates by performing a Monte Carlo nuclear reaction network calculation that reproduced all of the observed abundances in NGC 2419, including the Mg--K anticorrelation. They found a narrow band of possible temperature-density conditions between $\approx80-260$ MK and $\approx10^{-4}-10^{8}$ $\mathrm{g/cm}^{3}$. The only known polluter candidates that can overlap with this band are classical novae and SAGB stars. Dermigny and Iliadis \cite{Dermigny2017} investigated the sensitivity of these temperature-density conditions on individual thermonuclear reaction rates. They found that $^{39}\mathrm{K}(p, \gamma)^{40}\mathrm{Ca}$, the key potassium-destroying reaction, could cause a dramatic shift in the acceptable conditions. If the $^{39}\mathrm{K}(p, \gamma)^{40}\mathrm{Ca}$ reaction rate systematically increased by as little as a factor of 5, the temperature-density band shows significantly reduced scatter and tends toward lower temperatures, constraining the polluter candidates.

The most recent $^{39}\mathrm{K}(p, \gamma)^{40}\mathrm{Ca}$ reaction rate evaluation by Longland \emph{et al.} \cite{Longland2018} has large uncertainties between about 50 MK and 200 MK, coinciding with the range of acceptable temperatures found by Iliadis \emph{et al.} \cite{Iliadis2016}. This is particularly apparent at 80 MK, where the $1\sigma$ reaction rate uncertainty band spans almost a factor of 100. To constrain the temperature-density conditions of polluter stars in NGC 2419, it is therefore crucial to reduce the uncertainty in the $^{39}\mathrm{K}(p, \gamma)^{40}\mathrm{Ca}$ reaction rate.

The relevant temperature range of $80-260$ MK spans $^{39}\mathrm{K}+p$ resonance energies of about $E^{\mathrm{c.m.}}_{r} \approx 80-550$ keV ($E_{x} \approx 8400-8900$ keV), where the $^{40}\mathrm{Ca}$ proton separation energy is $S_{p} = 8328.18(2)$ keV \cite{Wang2021}. Given the technical challenges of measuring small cross sections at low energies, no direct $^{39}\mathrm{K}(p, \gamma)^{40}\mathrm{Ca}$ measurement \cite{Kikstra1990,Cheng1981,Leenhouts1966} has determined resonance strengths for resonances below $E^{\mathrm{c.m.}}_{r} = 606$ keV ($E_{x} = 8935$ keV), and indirect measurements \cite{Erskine1966,Seth1967,Forster1970,Cage1971,Fuchs1969} have suffered from low resolution, only obtaining spectroscopic factors for 4 unbound states below $E_{x} = 8935$ keV. In this work, we investigate the states in this region at a higher resolution than was previously attainable. Using the $^{39}\mathrm{K}(^{3}\mathrm{He}, d)^{40}\mathrm{Ca}$ proton-transfer reaction, we measure spectroscopic factors, proton partial-widths and resonance energies of 8 unbound $^{40}\mathrm{Ca}$ states below $E_{x} = 8935$ keV, including the 4 previously-reported states. We thus resolve 4 resonances in $^{39}\mathrm{K}+p$ for the first time.

%%%%%%%%%%%%%%%%%%%%%%%%%%%%%%%%%%%%%%%%%%%%%%%%%%%%%%%%%%%%%%%%%%%%%%%%%%%%%%%%%%%%%%%%
%\section{Experimental Methods}

The $^{39}\mathrm{K}(^{3}\mathrm{He}, d)^{40}\mathrm{Ca}$ experiment was performed using the Enge Split-Pole Spectrograph at the Triangle Universities Nuclear Laboratory (TUNL). The 10 MV FN tandem Van de Graaff accelerator at TUNL accelerated a fully-ionized $^{3}$He beam to 21 MeV, where the energy was stabilized using a pair of high-resolution slits between two $90^{\circ}$ dipole magnets. The target was produced by evaporating approximately 75 $\mu\mathrm{g/cm}^{2}$ of natural KI onto a 21 $\mu\mathrm{g/cm}^{2}$ natural-abundance C foil. The $^{39}\mathrm{K}(^{3}\mathrm{He}, d)^{40}\mathrm{Ca}$ and $^{39}\mathrm{K}(^{3}\mathrm{He}, {}^{3}\mathrm{He})^{39}\mathrm{K}$ reactions were measured between $\theta_{\mathrm{lab}}=5-20^{\circ}$ and $\theta_{\mathrm{lab}}=15-59^{\circ}$, respectively, using a focal-plane detector package \cite{Marshall2019} consisting of two position-sensitive avalanche counters, a $\Delta\mathrm{E}$ proportionality counter, and an E scintillator.

To minimize the effects of non-uniformity in the target, uncertainty in the target thickness, and target degradation after exposure to the beam, the $^{39}\mathrm{K}(^{3}\mathrm{He}, {}^{3}\mathrm{He})^{39}\mathrm{K}$ and $^{39}\mathrm{K}(^{3}\mathrm{He}, d)^{40}\mathrm{Ca}$ yields from the focal-plane detector were normalized to the simultaneous $^{39}\mathrm{K}(^{3}\mathrm{He}, {}^{3}\mathrm{He})^{39}\mathrm{K}$ yield of a Si detector telescope positioned at a constant $\theta_{\mathrm{lab}} = 45^{\circ}$ inside the target chamber. Normalizing to a global $^{3}\mathrm{He}$ optical model potential (OMP) for $^{39}\mathrm{K}$ then corrected the overall scale for the $^{39}\mathrm{K}(^{3}\mathrm{He}, d)^{40}\mathrm{Ca}$ differential cross-section. While this technique removes uncertainties in target thickness, target degradation, and beam current, it does introduce model uncertainties, which are accounted for below. More details on this method can be found in Refs. \cite{Marshall2020,Marshall2023}.

Spectroscopic factors $C^{2}S_{{}^{39}\mathrm{K}+p}$, or simply $C^{2}S$, are calculated by normalizing the $^{39}\mathrm{K}(^{3}\mathrm{He}, d)^{40}\mathrm{Ca}$ differential cross-section $d\sigma / d\Omega_{\mathrm{Exp}}$ to that of an appropriate distorted-wave Born approximation (DWBA) model $d\sigma / d\Omega_{\mathrm{DWBA}}$ via
\begin{equation*}
\frac{d\sigma}{d\Omega}_{\mathrm{Exp}} = C^{2}S_{{}^{39}\mathrm{K}+p} \, C^{2}S_{d+p} \, \frac{d\sigma}{d\Omega}_{\mathrm{DWBA}},
\end{equation*}
where we adopt $C^{2}S_{d+p} = 1.32(1)$ from \emph{ab initio} calculations \cite{Brida11}. The DWBA models used in the present work are calculated from the nuclear reaction code, \texttt{FRESCO} \cite{Thompson1988,FRESCO}, using the zero-range (ZR) approximation with a ZR coupling constant $D_{0} = - 172.8$ $\mathrm{MeV} \, \mathrm{F}^{3/2}$ from Ref. \cite{Bassel1966}. The proton partial-widths $\Gamma_{p}$ are calculated using dimensionless single-particle reduced widths from \texttt{FRESCO} bound-state radial wavefunctions evaluated at $R = 1.25 \, ( A_{t}^{1/3} + A_{p}^{1/3} )$ fm.

%%%%%%%%%%%%%%%%%%%%%%%%%%%%%%%%%%%%%%%%%%%%%%%%%%%%%%%%%%%%%%%%%%%%%%%%%%%%%%%%%%%%%%%%
%\subsection{Global Optical Model Potentials}

We use the global $^{3}\mathrm{He}$ OMP of Liang \emph{et al.} \cite{Liang2009} for the entrance channel and the global $d$ OMP of An and Cai \cite{An2006} for the exit channel in the calculation of DWBA cross-sections. Both potentials take an identical form, with the parameters given in Table \ref{tab:OMP}, including the proton bound-state parameters. The choice of parameters introduces systematic uncertainties into the calculation of $C^{2}S$ and $\Gamma_{p}$. We assign a fractional systematic uncertainty of $30\%$ for both quantities, which is typical of $(^{3}\mathrm{He}, d)$ reactions \cite{Endt1977}.

\begin{table*}[t]
\caption{\label{tab:OMP}Global optical model potential parameters for $^{39}\mathrm{K} + \,^{3}\mathrm{He}$ and $^{40}\mathrm{Ca} + d$ and the proton bound-state parameters.}
\begin{ruledtabular}
\begin{tabular}{cllllllllllllll}
 $\mathrm{particle}$ & $V_{r}$ & $r_{r}$ & $a_{r}$ & $W_{v}$ & $r_{v}$ & $a_{v}$ & $W_{s}$ & $r_{s}$ & $a_{s}$ & $V_{so}$ & $r_{so}$ & $a_{so}$ & $W_{so}$ & $r_{c}$\\ \midrule
 $^{3}\mathrm{He}$ & 117.881 & 1.178 & 0.768 & -0.646 & 1.415 & 0.847 & 20.665 & 1.198 & 0.852 & 2.083 & 0.738 & 0.946 & -1.159 & 1.289\\
 $d$ & 91.120 & 1.150 & 0.762 & 2.234 & 1.334 & 0.513 & 10.274 & 1.378 & 0.743 & 3.557 & 0.972 & 1.011 & & 1.303\\
 $p$ & $V_{p}\footnotemark[1]$ & 1.250 & 0.650 &  &  &  &  &  &  &  &  &  &  & 1.250\\
\end{tabular}
\end{ruledtabular}
\footnotetext[1]{Well depth is adjusted to reproduce the proton binding energy.}
\end{table*}

%%%%%%%%%%%%%%%%%%%%%%%%%%%%%%%%%%%%%%%%%%%%%%%%%%%%%%%%%%%%%%%%%%%%%%%%%%%%%%%%%%%%%%%%
%\subsection{New States Observed}

The high resolution of the Enge Split-Pole Spectrograph has enabled observation of states not resolved in the previous $^{39}\mathrm{K}(^{3}\mathrm{He}, d)^{40}\mathrm{Ca}$ proton-transfer measurements of Erskine \cite{Erskine1966}, Seth \emph{et al.} \cite{Seth1967}, Forster \emph{et al.} \cite{Forster1970}, and Cage \emph{et al.} \cite{Cage1971}. In those measurements, $C^{2}S$ was only calculated for 2 unbound $^{40}\mathrm{Ca}$ states, 8425 keV and 8551 keV. Similarly, the $^{39}\mathrm{K}(d, n)^{40}\mathrm{Ca}$ proton-transfer measurement of Fuchs \emph{et al.} \cite{Fuchs1969} calculated $C^{2}S$ for 4 unbound $^{40}\mathrm{Ca}$ states below 8935 keV, namely, 8359 keV, 8425 keV, 8551 keV, and 8665 keV. We calculate $C^{2}S$ and $\Gamma_{p}$ for 8 unbound states below 8935 keV in this work, including 4 states not resolved in the previous $^{39}\mathrm{K}(^{3}\mathrm{He}, d)^{40}\mathrm{Ca}$ and $^{39}\mathrm{K}(d, n)^{40}\mathrm{Ca}$ measurements.

\begin{figure}[t]
\includegraphics[width=8.6cm]{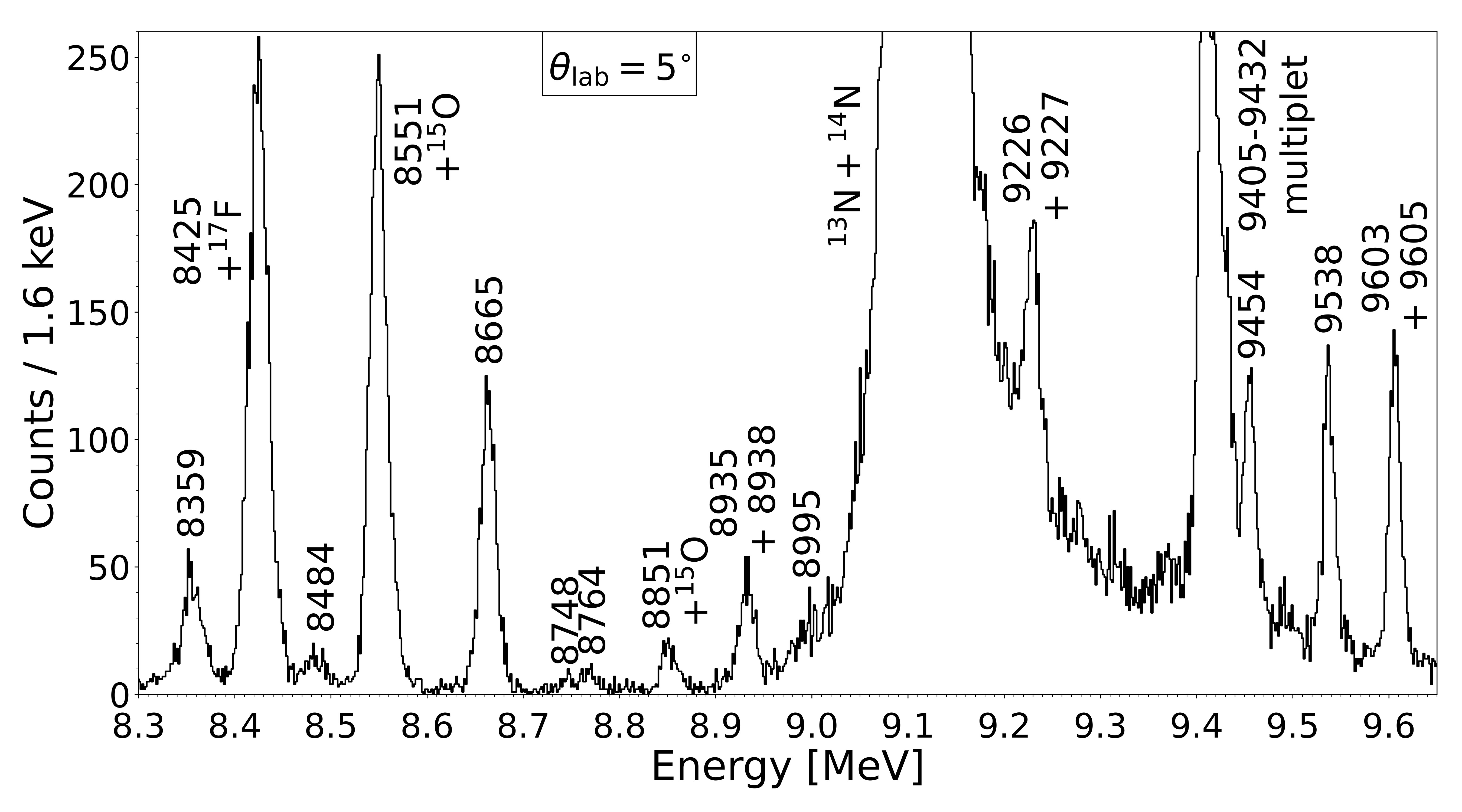} % 8.6 cm is exact column width. Height scaled automatically to keep aspect ratio
\caption{\label{fig:spectrum}Energy-calibrated focal-plane position spectrum gated on deuterons from the $^{39}\mathrm{K}(^{3}\mathrm{He}, d)^{40}\mathrm{Ca}$ reaction at $\theta_{\mathrm{lab}} = 5^{\circ}$ with a 21 MeV $^{3}\mathrm{He}$ beam. The unbound $^{40}\mathrm{Ca}$ states populated in the present experiment are labeled with their ENSDF \cite{Chen2017} energies in keV. This spectrum was collected using about 230 pnA of $^3$He$^{++}$ for 4.6 hours.}
\end{figure}

\begin{figure*}[t]
\centering
\includegraphics[width=17.2cm]{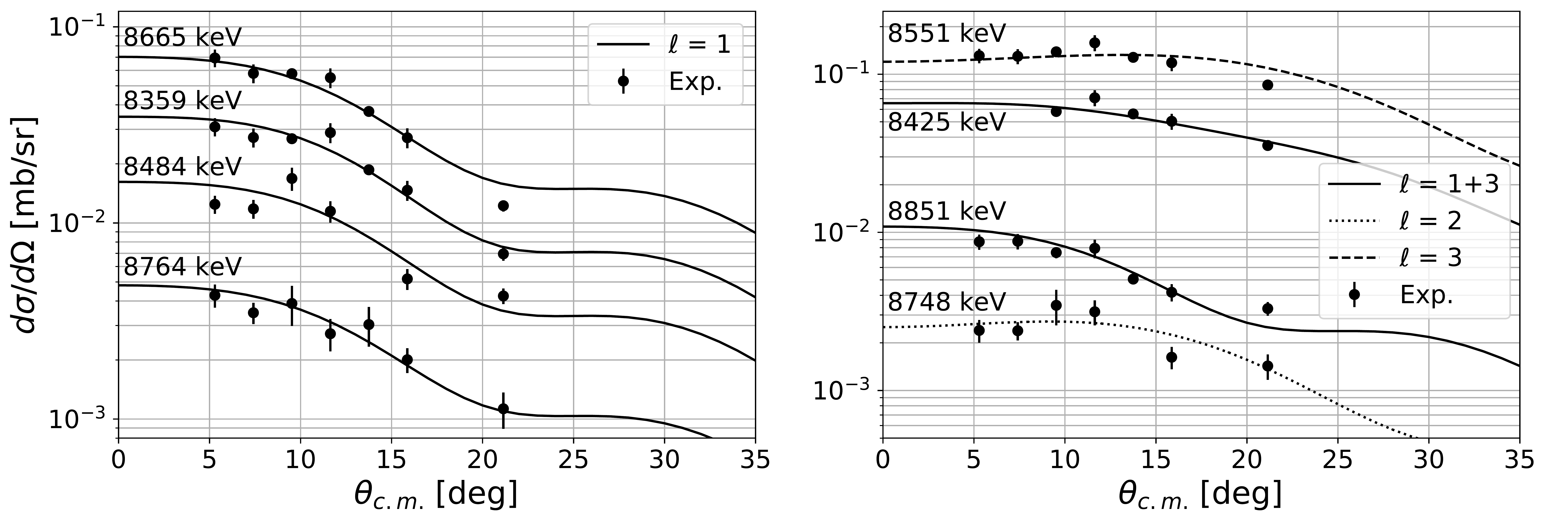}
\caption{\label{fig:unbound_l}Differential cross-sections of unbound $^{40}\mathrm{Ca}$ states below 8935 keV observed in the present experiment. The left panel shows the $l=1$ distributions, while the right panel shows all other distributions. The experimental data were normalized to the Si detector telescope yield, as described in the text. The zero-range DWBA model curves were computed using the nuclear reaction code, \texttt{FRESCO} \cite{Thompson1988,FRESCO}. Error bars represent statistical uncertainties only.}
\end{figure*}

Figure \ref{fig:spectrum} shows the unbound $^{40}\mathrm{Ca}$ states populated in the present experiment from an energy-calibrated focal-plane position spectrum gated on deuterons at $\theta_{\mathrm{lab}} = 5^{\circ}$. The $^{40}\mathrm{Ca}$ states are labeled with their ENSDF \cite{Chen2017} energies in keV, and contaminants from the target are also shown (namely $^{12}$C, $^{13}$C, $^{14}$N, and $^{16}$O). Due to their different kinematic properties, deuterons from these contaminants move across the focal plane, causing them to occlude peaks of interest at some angles, but also making them trivial to identify. The 4 previously-unresolved states correspond to ENSDF energies of 8484 keV, 8748 keV, 8764 keV, and 8851 keV. Energies were calibrated using a Markov Chain Monte Carlo method with the calibration states 4491 keV, 5614 keV, 6025 keV, 7532 keV, 8425 keV, and 9454 keV. The states 8484 keV and 9136 keV were used as replacement calibration points when 8425 keV and 9454 keV were obscured at $\theta_{\mathrm{lab}} = 5, 7^{\circ}$ and $\theta_{\mathrm{lab}} = 20^{\circ}$, respectively. Angular distributions of all 8 unbound $^{40}\mathrm{Ca}$ states below 8935 keV are shown in Figure \ref{fig:unbound_l}, along with the corresponding ZR DWBA models used to calculate $C^{2}S$ and $\Gamma_{p}$ for each state. We assume single-particle configurations of $2p_{3/2}$, $2d_{5/2}$, and $1f_{7/2}$ for all $l=1, \,2, \, \mathrm{and} \,\, 3$ distributions, respectively.

%%%%%%%%%%%%%%%%%%%%%%%%%%%%%%%%%%%%%%%%%%%%%%%%%%%%%%%%%%%%%%%%%%%%%%%%%%%%%%%%%%%%%%%%
%\subsection{Results}

Excitation energies, resonance energies, $J^{\pi}$ and $l$ assignments, spectroscopic factors, and proton partial-widths from this experiment are reported in Table \ref{tab:partialwidth}. The excitation energy of each state is adopted from ENSDF \cite{Chen2017}, except for the 8425 keV and 8748 keV states, where we instead report the more recent nuclear resonance fluorescence (NRF) energy measurement of Gribble \emph{et al.} \cite{Gribble2022}. Resonance energies are calculated with $S_{p} = 8328.18(2)$ keV from Ref. \cite{Wang2021}. Proton partial-width upper limits calculated by Longland \emph{et al.} \cite{Longland2018} are shown in the last column. We will now briefly discuss each unbound $^{40}\mathrm{Ca}$ state below $E_{x} = 8935$ keV observed in the present experiment.

\begin{table*}
\caption{\label{tab:partialwidth}Energies, angular momentum assignments, spectroscopic factors, and proton partial-widths obtained for unbound $^{40}\mathrm{Ca}$ states below $E_{x} = 8935$ keV in this experiment. Spectroscopic factors are reported with their statistical uncertainties only, while proton partial-widths are reported with their total uncertainty, including a $30\%$ systematic uncertainty. The last column shows proton partial-width upper limits calculated by Longland \emph{et al.} \cite{Longland2018}.}
\begin{ruledtabular}
\begin{tabular}{lllllllll}
$E_{x}$\footnotemark[1] [keV]&$E_{x}$\footnotemark[2] [keV]&$E_{r}$\footnotemark[2] [keV]&$J^{\pi}$\footnotemark[1]&$J^{\pi}$\footnotemark[2]&$l$\footnotemark[2]&$(2J+1)C^{2}S$\footnotemark[2]&$(2J+1)\Gamma_{p}$\footnotemark[2] [eV]&$(2J+1)\Gamma_{p, UL}$ [eV]\\ \midrule
8358.9(6)&8357.3(25)&29.1(25)&$(0,1,2)^{-}$&$(0,1,2,3)^{-}$&1&0.191(11)&$2.62(80) \times 10^{-38}$&$5.88 \times 10^{-36}$\\
8424.35(31)\footnotemark[3]\footnotemark[4]& \emph{8425}\footnotemark[4] & \emph{96.17(31)}\footnotemark[3] &$2^{-}$&$(1,2,3)^{-}$&$1 + 3$&$0.185(29) \phantom{a}(l=1)$&$5.4(18) \times 10^{-17}$&$1.47 \times 10^{-15}$\\
 & & & & & &$1.23(10) \phantom{aa}(l=3)$&$3.9(12) \times 10^{-19}$& \\
8484.02(13)\footnotemark[4]&8482.3(29)&154.1(29)&$(1^{-},2^{-},3^{-})$&$(0,1,2,3)^{-}$&1&0.093(6)&$8.1(25) \times 10^{-12}$&$1.16 \times 10^{-9}$\\
8551.1(7)&8549.5(25)&221.3(25)&$5^{-}$&$(1-5)^{-}$&3&4.58(25)&$1.33(40) \times 10^{-9}$&$6.60 \times 10^{-9}$\\
8665.3(8)&8662.9(26)&334.7(26)&$1^{-}$&$(0,1,2,3)^{-}$&1&0.444(24)&$1.56(48) \times 10^{-4}$&$1.10 \times 10^{-3}$\\
8748.59(19)\footnotemark[3]&8743.6(28)&415.4(28)&$2^{+}$&$(0-4)^{+}$&2&0.020(2)&$1.82(57) \times 10^{-5}$&$1.81 \times 10^{-1}$\\
8764.18(6)&8767.3(27)&439.1(27)&$3^{-}$&$(0,1,2,3)^{-}$&1&0.035(3)&$7.1(22) \times 10^{-4}$&$1.20 \times 10^{-1}$\\
8850.6(9)&8849.4(28)&521.2(28)&$6^{-},7^{-},8^{-}$&$(1,2,3)^{-}$&$1 + 3$&$0.055(3) \phantom{aa}(l=1)$&$1.10(34) \times 10^{-2}$&$1.86 \times 10^{-4}$\\
 & & & & & &$0.055(3) \phantom{aa}(l=3)$&$2.13(65) \times 10^{-5}$& \\
\end{tabular}
\end{ruledtabular}
\footnotetext[1]{From ENSDF \cite{Chen2017}, unless otherwise indicated.}
\footnotetext[2]{From the present experiment, unless otherwise indicated.}
\footnotetext[3]{From Gribble \emph{et al.} \cite{Gribble2022}.}
\footnotetext[4]{Level used as a calibration point for at least one angle in this work, or for all unobscured angles in the case of 8425 keV. See text.}
\end{table*}

\emph{$\mathbf{E_{x} = 8359}$} \textbf{keV}; \emph{$\mathbf{E^{\boldsymbol{\mathrm{c.m.}}}_{r} = 29}$} \textbf{keV:} \, The spin-parity of this state is assigned $J^{\pi} = (0, 1, 2)^{-}$ in ENSDF based on $(d, n)$ \cite{Fuchs1969} and $(p, p'\gamma)$ \cite{Tellez1973}. The present work assigns $l=1$, in agreement with Ref. \cite{Fuchs1969}. Longland \emph{et al.} \cite{Longland2018} did not use the $(d,n)$ \cite{Fuchs1969} $C^{2}S$ measurement in their $^{39}\mathrm{K}(p, \gamma)^{40}\mathrm{Ca}$ reaction rate calculation because of the 12 keV energy discrepancy with ENSDF. This state was instead assigned a proton partial-width upper limit which the present work now replaces.

\emph{$\mathbf{E_{x} = 8425}$} \textbf{keV}; \emph{$\mathbf{E^{\boldsymbol{\mathrm{c.m.}}}_{r} = 96}$} \textbf{keV:} \, The spin-parity of this state is assigned $J^{\pi} = 2^{-}$ in ENSDF based on $(d, n)$ \cite{Fuchs1969}, $(^{3}\mathrm{He}, \alpha)$ \cite{Cline1974}, $(p, p')$ \cite{Horen1985,Hosono1982}, $(e, e')$ \cite{Steffen1980}, and $(^{3}\mathrm{He}, d)$ \cite{Erskine1966}. It has also since been confirmed via NRF \cite{Gribble2022}. The present work assigns $l=1+3$, in agreement with $(d, n)$ \cite{Fuchs1969}, but in disagreement with $(^{3}\mathrm{He}, d)$ \cite{Erskine1966,Seth1967,Cage1971}, where $l=3$ is assigned. Longland \emph{et al.} \cite{Longland2018} uses the $(^{3}\mathrm{He},d)$ $C^{2}S$ measurement of Ref. \cite{Cage1971} in their analysis.

\emph{$\mathbf{E_{x} = 8484}$} \textbf{keV}; \emph{$\mathbf{E^{\boldsymbol{\mathrm{c.m.}}}_{r} = 154}$} \textbf{keV:} \, This weakly-populated state has not been observed in any previous $(^{3}\mathrm{He}, d)$ or $(d, n)$ measurement. Its spin-parity is assigned $J^{\pi} = (1^{-},2^{-},3^{-})$ in ENSDF based on $(^{3}\mathrm{He}, \alpha)$ \cite{Cline1974}. The present work assigns $l=1$ in support of the ENSDF $J^{\pi}$ assignment. This measurement replaces the proton partial-width upper limit of Ref. \cite{Longland2018}.

\emph{$\mathbf{E_{x} = 8551}$} \textbf{keV}; \emph{$\mathbf{E^{\boldsymbol{\mathrm{c.m.}}}_{r} = 221}$} \textbf{keV:} \, The spin-parity of this state is assigned $J^{\pi} = 5^{-}$ in ENSDF based on $(^{3}\mathrm{He}, d)$ \cite{Erskine1966,Seth1967,Cage1971}, $(d, n)$ \cite{Fuchs1969}, $(p, t)$ \cite{Seth1977,Seth1974,Debevec1974}, and $(p, p')$ \cite{Gruhn1972}. The present work assigns $l=3$, in agreement with the previous $(^{3}\mathrm{He}, d)$ and $(d, n)$ measurements.

\emph{$\mathbf{E_{x} = 8665}$} \textbf{keV}; \emph{$\mathbf{E^{\boldsymbol{\mathrm{c.m.}}}_{r} = 335}$} \textbf{keV:} \, The spin-parity of this state is assigned $J^{\pi} = 1^{-}$ in ENSDF based on $(d, n)$ \cite{Fuchs1969} and $(p, p'\gamma)$ \cite{Tellez1973}. The present work assigns $l=1$ in agreement with $(d, n)$.

\emph{$\mathbf{E_{x} = 8748}$} \textbf{keV}; \emph{$\mathbf{E^{\boldsymbol{\mathrm{c.m.}}}_{r} = 415}$} \textbf{keV:} \, This weakly-populated state was observed in $(d, n)$ \cite{Fuchs1969} as part of an unresolved doublet with the 8764 keV state, but no $l$ assignments were made. We have resolved these states in the present experiment. The spin-parity of the 8748 keV state is assigned $J^{\pi} = 2^{+}$ in ENSDF from $(p, p')$ \cite{Horen1984} and NRF \cite{Hartmann2002,Moreh1982} and has since been confirmed by an additional NRF experiment \cite{Gribble2022}. The present work assigns $l=2$ in support of the ENSDF $J^{\pi}$ assignment. This measurement replaces the proton partial-width upper limit of Ref. \cite{Longland2018}.

\emph{$\mathbf{E_{x} = 8764}$} \textbf{keV}; \emph{$\mathbf{E^{\boldsymbol{\mathrm{c.m.}}}_{r} = 439}$} \textbf{keV:} \, This state is part of a doublet with 8748 keV and was not resolved in $(d, n)$ \cite{Fuchs1969}. Its spin-parity is assigned $J^{\pi} = 3^{-}$ in ENSDF from $(p, t)$ \cite{Seth1977}. The present work assigns $l=1$, in support of the ENSDF $J^{\pi}$ assignment. This measurement replaces the proton partial-width upper limit of Ref. \cite{Longland2018}.

\emph{$\mathbf{E_{x} = 8851}$} \textbf{keV}; \emph{$\mathbf{E^{\boldsymbol{\mathrm{c.m.}}}_{r} = 521}$} \textbf{keV:} \, The spin-parity of this state is assigned $J^{\pi} = 6^{-}, 7^{-}, 8^{-}$ in ENSDF based on $(p, p')$ \cite{Gruhn1972}. It was observed in $(d, n)$ \cite{Fuchs1969}, but no $l$ assignment was made. The present experiment assigns $l=1+3$, suggesting $J^{\pi} = (1, 2, 3)^{-}$, in stark disagreement with ENSDF. The upper limit from Ref. \cite{Longland2018} for this state uses the ENSDF $J^{\pi}$ assignment. For this reason, the $(2J+1)\Gamma_{p}$ value from this work is about 2 orders of magnitude larger than the upper limit. Note that at $\theta_{\text{lab}}=5^{\circ}$ and $7^{\circ}$, this state is partially contaminated by the $E_x=7556$ keV state in $^{15}$O, which we estimate to contribute just $5\%$ of the total peak area.

%%%%%%%%%%%%%%%%%%%%%%%%%%%%%%%%%%%%%%%%%%%%%%%%%%%%%%%%%%%%%%%%%%%%%%%%%%%%%%%%%%%%%%%%
%\section{The $^{39}\mathrm{K}(p, \gamma)^{40}\mathrm{Ca}$ Reaction Rate}

\begin{figure}[t]
\includegraphics[width=8.6cm]{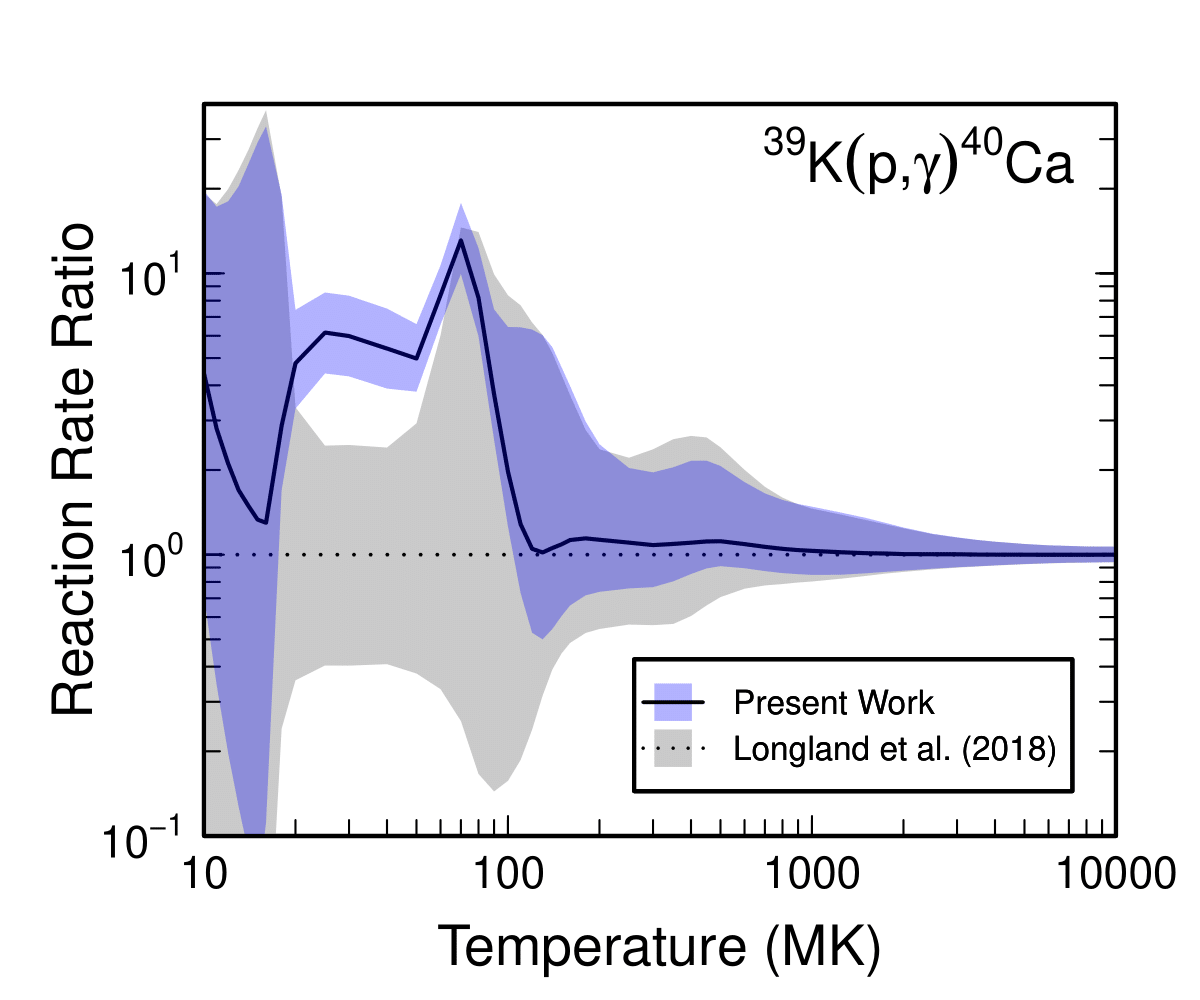} % 8.6 cm is exact column width. Height scaled automatically to keep aspect ratio
\caption{\label{fig:rateCompare}Comparison between the $^{39}\mathrm{K}(p, \gamma)^{40}\mathrm{Ca}$ reaction rate using the proton partial-widths and resonance energies of the present experiment (solid line, blue band) and the most recent evaluation of Longland \emph{et al.} \cite{Longland2018} (dotted line, gray band). The reaction rate ratio is taken with respect to the median, recommended rate of Ref. \cite{Longland2018} for both calculations. The $1\sigma$ uncertainty bands are shown.}
\end{figure}

The Monte Carlo reaction rate code, \texttt{RatesMC} \cite{Longland2010a,RatesMC}, was used to calculate a new $^{39}\mathrm{K}(p, \gamma)^{40}\mathrm{Ca}$ reaction rate probability density from the partial-widths and resonance energies reported in this work. Among the states observed in this experiment, only those that do not already have a directly measured resonance strength from $^{39}\mathrm{K}(p, \gamma)^{40}\mathrm{Ca}$ \cite{Kikstra1990,Cheng1981,Leenhouts1966}, i.e. only those below $E_{x} = 8935$ keV, were modified from the most recent reaction rate evaluation of Longland \emph{et al.} \cite{Longland2018}. The new reaction rate is compared with that of Ref. \cite{Longland2018} in Fig. \ref{fig:rateCompare}. The solid line and blue band represent the median, recommended rate and the $1\sigma$ uncertainty band of this work, respectively. The dotted line and gray band represent that of Ref. \cite{Longland2018}, except with resonance energies calculated using $S_{p} = 8328.18(2)$ keV from Wang \emph{et al.} \cite{Wang2021} for consistency, but with marginal effect. Both rates are normalized to the median, recommended rate of Ref. \cite{Longland2018}.

The large uncertainty in Ref. \cite{Longland2018} between about $50-200$ MK corresponds to most of the relevant temperatures that reproduce the Mg--K anticorrelation in the globular cluster NGC 2419 \cite{Iliadis2016}. Fig. \ref{fig:rateCompare} illustrates that the new reaction rate increases significantly below about 110 MK, up to a factor of 13 at 70 MK. The total width of the $1\sigma$ uncertainty band is also significantly reduced in this region, from a factor of 84 at 80 MK to just a factor of 2, a reduction of a factor of 42.

Our new resolution of the 154 keV resonance is primarily responsible for the increase in the rate and decrease in the uncertainty between about 55 MK and 110 MK. Similar effects occur between about 20 MK and 55 MK, primarily from our $l=1+3$ assignment of the 96 keV resonance, which has replaced the $l=3$ assignment of Ref. \cite{Cage1971} in this calculation. Note that this replacement has a negligible effect on the results mentioned above at 70 and 80 MK. Our resolution of the 29 keV resonance is responsible for the rate increase below 20 MK. The smaller effects above 110 MK result from a combination of the other resonances.

Nuclear reaction network calculations were used to investigate the effect the new rate has on potassium abundance in hydrogen-burning environments that reproduce the Mg--K anticorrelation observed in NGC 2419. We find that using a constant $T = 100$ MK and $\rho = 4 \times 10^{7}$ $\mathrm{g/cm}^{3}$ during hydrogen burning, our new rate reduces potassium abundance [K/Fe] by up to a factor of 1.7 compared to the rate of Ref. \cite{Longland2018}. Our new rate also reproduces the observed Mg--K anticorrelation for this $T-\rho$ condition, whereas the rate of Ref. \cite{Longland2018} produces too much potassium. This result suggests polluter candidates that burn hydrogen at lower temperatures, such as SAGB stars, may be more likely to reproduce the Mg--K anticorrelation in NGC 2419 than previously considered. A full phase space search, such as that of Iliadis \emph{et al.} \cite{Iliadis2016}, is beyond the scope of this work, but further investigation of the polluter candidates with the new reaction rate will be addressed in a future work.

%%%%%%%%%%%%%%%%%%%%%%%%%%%%%%%%%%%%%%%%%%%%%%%%%%%%%%%%%%%%%%%%%%%%%%%%%%%%%%%%%%%%%%%%
%\section{Conclusion}

We have reported new proton partial-widths for 4 unbound $^{40}\mathrm{Ca}$ states in the present work, corresponding to the 154 keV, 415 keV, 439 keV, and 521 keV resonances. Among these, the 154 keV resonance has by far the largest impact on the $^{39}\mathrm{K}(p, \gamma)^{40}\mathrm{Ca}$ reaction rate, increasing it by a factor of 13 at about 70 MK from the most recent evaluation of Longland \emph{et al.} \cite{Longland2018}. Another major impact on the reaction rate comes from our $l=1+3$ assignment of the 96 keV resonance. The significant increase in the reaction rate below 110 MK from these resonances leads to a reduction in potassium abundance by as much as a factor of 1.7 at $T = 100$ MK and $\rho = 4 \times 10^{7}$ $\mathrm{g/cm}^{3}$ in nuclear reaction network calculations. This result opens up the possibility for polluter candidates that burn hydrogen at lower temperatures to reproduce the Mg--K anticorrelation in NGC 2419. To further constrain the hydrogen-burning conditions, continued focus should be on (i) constraining the resonance strength of the 335 keV resonance, which still dominates the reaction rate between about $150-500$ MK; and (ii) constraining the other key reactions, namely $^{30}$Si(p,$\gamma$)$^{31}$P, $^{37}$Ar(p,$\gamma$)$^{38}$K, and $^{38}$Ar(p,$\gamma$)$^{39}$K.

%%%%%%%%%%%%%%%%%%%%%%%%%%%%%%%%%%%%%%%%%%%%%%%%%%%%%%%%%%%%%%%%%%%%%%%%%%%%%%%%%%%%%%%%
%%%%%%%%%%%%%%%%%%%%%%%%%%%%%%%%%%%%%%%%%%%%%%%%%%%%%%%%%%%%%%%%%%%%%%%%%%%%%%%%%%%%%%%%
\begin{acknowledgments}
We would like to thank Kiana Setoodehnia for being crucial to the success of the experimental program involved with the TUNL Enge Split-Pole Spectrograph. We also greatly appreciate many helpful comments from Thanassis Psaltis. This material is based partly upon work supported by the U.S. Department of Energy, Office of Science, Office of Nuclear Physics, under Award Number DE-SC0017799 and Contract Nos. DE-FG02-97ER41033 and DE-FG02-97ER41042.
\end{acknowledgments}

\end{document}